\documentstyle[12pt,cite,graphics,axodraw]{article}
\newcommand\Rsep{R_{\mathrm{sep}}}
\newcommand\as{\alpha_{\mathrm{s}}}
\newcommand\kt{{k_\perp}}
\newcommand\ycut{y_{\mathrm{cut}}}
\newcommand\smfrac[2]{{\textstyle\frac{#1}{#2}}}

\begin{document}
\topskip 2cm
\begin{titlepage}

\renewcommand{\thefootnote}{\fnsymbol{footnote}}
\begin{center}
{\large\bf JET PHENOMENOLOGY\footnote{Contributed to Proceedings of Les
    Rencontres de la Val\'ee d'Aoste: Results and Perspectives in
    Particle Physics, La Thuile, Italy, March 2--8, 1997.}} \\
\vspace{2.5cm}
{\large Michael H. Seymour} \\
\vspace{.5cm}
{\sl Rutherford Appleton Laboratory, Chilton,}\\
{\sl Didcot, Oxfordshire.  OX11 0QX.  U.K.}\\
\vspace{2.5cm}
\vfil
\begin{abstract}
  We discuss the phenomenology of jet physics at hadron colliders,
  concentrating on the internal structure of jets, which is studied
  using the jet shape distribution or subjet distributions.
\end{abstract}

\end{center}
\end{titlepage}
\renewcommand{\thefootnote}{\arabic{footnote}}
\setcounter{footnote}{0}

\section{Introduction}

The inclusive jet rate in hadron collisions appears to be pretty well
understood both experimentally\cite{Phil,CDF1,D01} and
theoretically\cite{ACGG,EKS1,GGK1}.  The data are in good agreement with
theory over seven orders of magnitude in rate, although with a hint of
an excess at high $E_T$.  There is little dependence on parton
distribution functions and the overall theoretical error is estimated to
be small.


However, one of the cross-checks often used to ensure that the jet data
are well understood and modelled is the jet shape: a simple measure of
the internal structure of the jets, specifically of how broad they are.
Here the picture is not so clear\cite{Terry,CDF3,D03}.  The experiments
are in good agreement with each other, giving us confidence in the data,
but the experimental jets are considerably broader than predicted by
hadron-level Monte Carlo event generators, and the dependence on the jet
rapidity is not well modelled either.  Next-to-leading order (NLO)
calculations of the jet rate give a leading order (LO) prediction for
the internal jet structure\cite{EKS3,GGK3}.  Although these can be tuned
to fit the data in any given bin in rapidity and transverse momentum,
they do not give a good prediction of the dependence on those
variables\cite{KK}.

In this paper, we briefly summarize a recent study\cite{paper} of the
extent to which these LO predictions should be affected by higher
orders, resummation of logarithmically enhanced terms to all orders, and
by including non-perturbative hadronization corrections.  The conclusion
is that all these effects are important, and we should not be surprised
if LO does not agree well with data.  As part of the study, we asked how
well suited currently-used jet algorithms are for quantitatively probing
the internal structure of jets.  We found that the iterative cone
algorithm\cite{CDF4}, used by essentially all current experiments, is
not infrared safe beyond the three-parton final state (i.e.\ beyond NLO
for inclusive jet and dijet cross sections, but beyond LO for internal
jet properties or three jet cross sections\cite{GK}).  This makes them
hopeless for quantitative studies.  This problem can be solved by a
slight modification to the algorithm\cite{LdP}, or better still by
abandoning cone-type algorithms altogether in favour of the cluster-type
$\kt$ algorithm\cite{CDSW,ES}.

In Sect.~2, we describe the jet definitions in current use.  Since each
experiment defines their own slightly different variant of the cone
algorithm, we concentrate on one in particular, D\O's\cite{D04}, and only
indicate the differences with respect to other experiments' where
relevant.  We then calculate, in a simple approximation, the cross
section to next-to-next-to-leading order (NNLO) according to this
algorithm, and explicitly show that it is not infrared safe.  We discuss
the solution proposed in \cite{LdP}.

In Sect.~3, we calculate the LO predictions for the jet shape in the
various algorithms we have discussed.  We estimate the effect of NLO
corrections, power-suppressed hadronization corrections, and resummation
of large logarithms to all orders.

In Sect.~4, we discuss another way of probing the internal structure of
jets: by resolving subjets within them.  This has many advantages over
the jet shape, not least the fact that the subjet resolution variable
gives us an extra handle to turn.  We can choose to sit in a very
perturbative regime, or to move smoothly into the hadronization regime,
and eventually for very small resolution parameters, obtain the results
of the usual jet shape as a limit of the subjet study.

Finally in Sect.~5, we make some concluding remarks.

\section{Jet definitions and cross sections}

All the algorithms we discuss define the momentum of a jet in terms of
the momenta of its constituent particles in the same way, inspired by
the Snowmass accord\cite{Snowmass}.  The transverse energy, $E_T$,
pseudorapidity, $\eta$, and azimuth, $\phi$, are given by:
\begin{eqnarray}
\label{snowmass}
  E_{T\mathrm{jet}} &=& \sum_{i\in\mathrm{jet}} E_{Ti},\nonumber\\
  \eta_{\mathrm{jet}} &=& \sum_{i\in\mathrm{jet}}
  E_{Ti}\,\eta_{i}/E_{T\mathrm{jet}},\\
  \phi_{\mathrm{jet}} &=& \sum_{i\in\mathrm{jet}}
  E_{Ti}\,\phi_{i}/E_{T\mathrm{jet}}.\nonumber
\end{eqnarray}
We shall always use boost-invariant variables, so whenever we say
`angle', we mean the Lorentz-invariant opening angle $R_{ij} =
\sqrt{(\eta_i-\eta_j)^2 + (\phi_i-\phi_j)^2}$.  Also, whenever we say
`energy', we mean transverse energy, $E_T=E\sin\theta$.

\subsection{The \boldmath$\kt$ algorithm}

We discuss the fully-inclusive $\kt$ algorithm including an $R$
parameter\cite{ES}.  It clusters particles (partons or calorimeter
cells) according to the following iterative steps:
\begin{enumerate}
\item
  For every pair of particles, define a closeness
\begin{equation}
  d_{ij} = \min(E_{Ti},E_{Tj})^2R_{ij}^2
  \left(\approx \min(E_i,E_j)^2\theta_{ij}^2 \approx k_\perp^2\right).
\end{equation}
\item
  For every particle, define a closeness to the beam particles,
\begin{equation}
  d_{ib} = E_{Ti}^2 R^2.
\end{equation}
\item
  If $\min\{d_{ij}\}<\min\{d_{ib}\},$ {\em merge\/} particles $i$ and
  $j$ according to Eq.~(\ref{snowmass}) (other merging schemes are also
  possible\cite{CDSW}).
\item
  If $\min\{d_{ib}\}<\min\{d_{ij}\},$ jet $i$ is {\em complete}.
\end{enumerate}
These steps are iterated until all jets are complete.  In this case, all
opening angles within each jet are $<R$ and all opening angles between
jets are $>R$.

\subsection{The D\O\ algorithm}

Since this is the main algorithm we shall study, we define it in full
detail.  It is based on the iterative-cone concept, with cone radius $R$.
Particles are clustered into jets according to the following steps:
\begin{enumerate}
\item The particles are passed through a calorimeter with cell size
  $\delta_0\times\delta_0$ in $\eta\times\phi$ (in D\O, $\delta_0=0.1$).
  In the parton-level algorithm, we simulate this by clustering together
  all partons within an angle $\delta_0$ of each other.
\item Every calorimeter cell (cluster) with energy above $E_0$, is
  considered as a `seed cell' for the following step (in D\O,
  $E_0=1$~GeV).
\item\label{reiterate} A jet is defined by summing all cells within an
  angle $R$ of the seed cell according to Eq.~(\ref{snowmass}).
\item If the jet direction does not coincide with the seed cell,
  step~\ref{reiterate} is reiterated, replacing the seed cell by the
  current jet direction, until a stable jet direction is achieved.
\item We now have a long list of jets, one for each seed cell.  Many are
  duplicates: these are thrown away\footnote{In D\O, any with energy
    below 8~GeV are also thrown away.  For jets above 16~GeV, this makes
    only a small numerical difference, which is not important to our
    discussion, so we keep them.}.
\item\label{merge} Some jets could be overlapping.  Any jet that has
  more than 50\% of its energy in common with a higher-energy jet is
  merged with that jet: all the cells in the lower-energy jet are
  considered part of the higher-energy jet, whose direction is again
  recalculated according to Eq.~(\ref{snowmass}).
\item\label{split} Any jet that has less than 50\% of its energy in
  common with a higher-energy jet is split from that jet: each cell is
  considered part only of the jet to which it is nearest.
\end{enumerate}
Note that despite the use of a fixed cone of radius $R$, jets can
contain energy at angles greater than $R$ from their direction, because
of step~\ref{merge}.  This is not a particular problem.  This is
essentially also the algorithm used by ZEUS (PUCELL), except that their
merging/splitting threshold is 75\% instead of 50\%.  The CDF algorithm
is similar again, and also uses 75\%, but has a slightly different
splitting procedure.

\subsection{Jet cross sections}

\begin{figure}
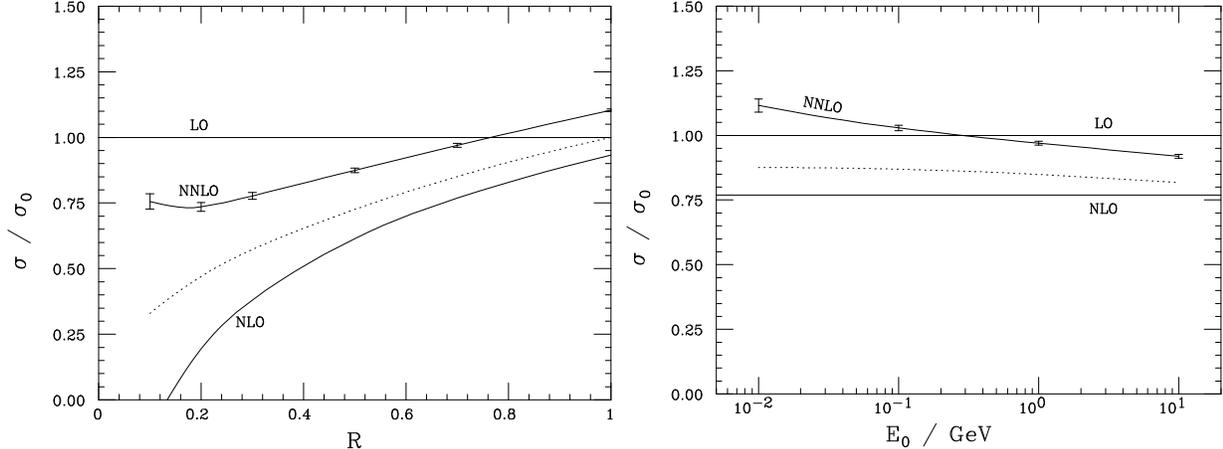

  \centerline{
    \resizebox{!}{6cm}{\includegraphics{profile_01.ps}}
    \hfill
    \resizebox{!}{6cm}{\includegraphics{profile_02.ps}}
    }
  \caption[]{{\it The radius dependence with $E_0=1$~GeV (left) and seed
      cell threshold dependence with $R=0.7$ (right) of the inclusive
      jet cross section in the D\O\ jet algorithm in fixed-order (solid)
      and all-orders (dotted) calculations.  The error bars come from
      Monte Carlo statistics.}}
  \label{itxsec}
\vspace*{-2ex}
\end{figure}

The issue of infrared safety in jet cross sections is discussed in
\cite{paper}.  There we define a class of jet definitions that we call
`almost unsafe', in which the definition appears at first sight to be
unsafe, but that some minor detail makes it safe, although still
unreliable.  The iterative cone algorithm is of exactly this type, as
can be seen in Fig.~\ref{itxsec}, where we show the cross section at
successive orders, in the double-logarithmic approximation.
The NNLO cross section depends logarithmically on the energy threshold
of a calorimeter cell, and for an ideal calorimeter with zero threshold
is clearly infinite.  This divergence comes from the separation of
events on the two-/three-jet boundary, and corresponds exactly to the
divergence to negative infinity found for the three-jet cross section in
\cite{GK}.

\begin{figure}
\centerline{
  \begin{picture}(252,144)(0,0)
    \SetWidth{1.5}
    \LongArrow(126,0)(72,108)
    \LongArrow(126,0)(180,108)
    \Oval(72,108)(36,72)(0)
    \Oval(180,108)(36,72)(0)
    \Text(63,21)[c]{\large(a)}
  \end{picture}
~\hfill~
  \begin{picture}(252,144)(0,0)
    \SetWidth{1.5}
    \LongArrow(126,0)(72,108)
    \LongArrow(126,0)(180,108)
    \Oval(72,108)(36,72)(0)
    \Oval(180,108)(36,72)(0)
    \SetWidth{0.5}
    \Photon(126,36)(126,104){3}{4}
    \LongArrow(126,104)(126,108)
    \SetWidth{0.5}
    \Oval(126,108)(36,72)(0)
    \Text(63,21)[c]{\large(b)}
  \end{picture}
}
  \caption[]{{\it Illustration of the problem region for the iterative cone
      algorithm.  In (a), there are two hard partons, with overlapping
      cones.  In (b) there is an additional soft parton in the overlap
      region.}}
  \label{ill}
\vspace*{-2ex}
\end{figure}
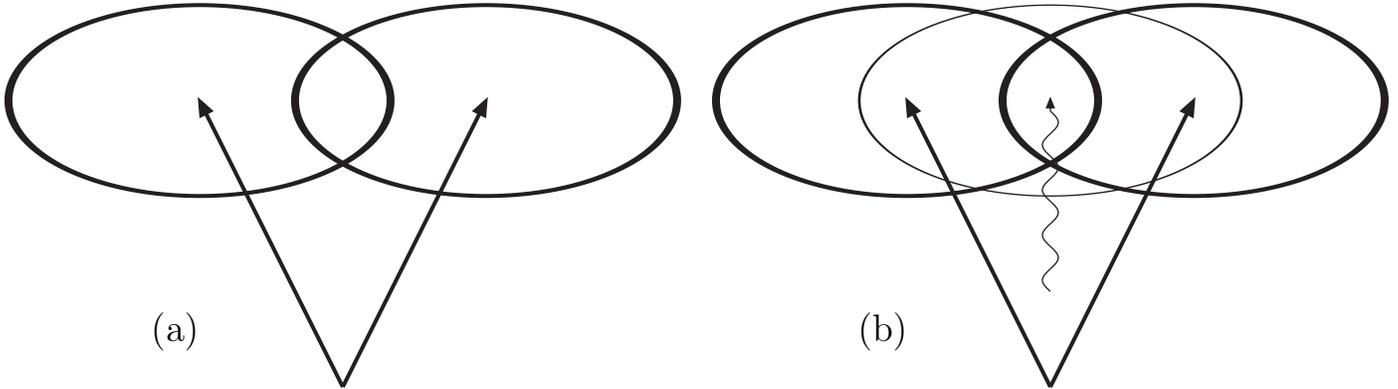

The strong $E_0$ dependence can be easily understood.  It arises from
configurations that have long been understood to be a problem in cone
algorithms, where two partons lie somewhere between $R$ and $2R$ apart
in angle, but sufficiently balanced in energy that they are both within
$R$ of their common centre, defined by Eq.~(\ref{snowmass}).  This is
illustrated in Fig.~\ref{ill}a.
According to the iterative cone algorithm, each is a separate jet,
because the cone around each seed cell contains no other active cells,
so is immediately stable.  Although the two cones overlap, there is no
energy in the overlap region, so the splitting procedure is trivial,
and it is classed as a two-jet configuration.

Now consider almost the same event, but with the addition of a soft
parton, close to the energy threshold $E_0$, illustrated in
Fig.~\ref{ill}b.  If it is marginally below threshold, the
classification is as above, with the soft parton being merged with
whichever hard parton it is nearest.  If on the other hand it is
marginally above threshold, there is an additional seed cell.  The cone
around this seed encloses both the hard partons and thus a third stable
solution is reached.  Now the merging and splitting procedure produces
completely different results.  In either the CDF or D\O\ variants the
result is the same: each of the outer jets overlaps with the central
one, with the overlap region containing 100\% of the outer one's energy.
Thus each is merged with the central one, and it is classified as a
one-jet configuration.

The classification is different depending on whether or not there is a
parton in the overlap region with energy above $E_0$.  Since the
probability for this to occur can be estimated as
$\sim\frac{2C_A}{\pi}\as\log E_T/E_0$, the inclusive jet cross section
depends logarithmically on the energy threshold above which calorimeter
cells are considered seed cells.  Thus {\it the iterative cone jet
  definition is not fully infrared safe}.

It is worth recalling how the $\kt$ algorithm completely avoids this
issue, and remains infrared safe to all orders.  Merging starts with the
softest (lowest relative $\kt$) partons.  Thus in the configuration of
Fig.~\ref{ill}b, the soft parton is first merged with whichever hard
parton it is nearer.  Only then is any decision made about whether to
merge the two jets, based solely on their opening angle.  The algorithm
has completely `forgotten about' the soft parton, and treats the
configurations of Figs.~\ref{ill}a and~\ref{ill}b identically.  Thus,
details of the calorimeter's energy threshold become irrelevant,
provided it is significantly smaller than the jet's energy.

In Fig.~\ref{itxsec}, results were also shown from an all-orders
calculation in the same approximation.  As would be expected, the poor
behaviour at small $R$ has been tamed to give a well-behaved physical
prediction.  Surprisingly, the same is true of the $E_0$-dependence,
which is much milder in the all-orders result than in the NNLO result.

This can be understood as a Sudakov-type effect.  Although the fraction
of events with a hard emission in the problem region is small, the
probability of subsequent soft emission into the overlap of the cones in
those events is large, $\sim\frac{2C_A}{\pi}\as\log E_T/E_0\sim1$.  This
is precisely the logarithmic behaviour seen in the NNLO result of
Fig.~\ref{itxsec}.  However, when going to the all-orders result, the
probability of non-emission exponentiates, and we obtain
\begin{equation}
  \label{suda}
  \frac{2C_A}{\pi}\as\log E_T/E_0 \longrightarrow
  1-\exp\left(-\frac{2C_A}{\pi}\as\log E_T/E_0\right)
  = 1-\left(\frac{E_0}{E_T}\right)^{\frac{2C_A}{\pi}\as},
\end{equation}
the much slower behaviour seen in the all-orders result of
Fig.~\ref{itxsec}.

This result has a simple physical interpretation: in the `all-orders
environment', there are so many gluons around that there is almost
always at least one seed cell in the overlap region and the two jets are
merged to one.  In our simple approximation, the coupling
does not run.  If we retained the running coupling, this statement would
become even stronger, because the probability to emit soft gluons would
be even more enhanced.

It is precisely this effect that has lead to the belief that the merging
issue is a relatively unimportant numerical effect: \emph{in the
  experimental environment it is}.  However, expanding out the
exponential of Eq.~(\ref{suda}) as an order-by-order expansion in $\as$,
we obtain large coefficients at every order, and no hope of well-behaved
theoretical predictions.

Thus, \emph{if we are to study the internal properties of jets
  quantitatively, we must solve the overlap problem, to define jets in a
  perturbatively-calculable way.}

\begin{figure}
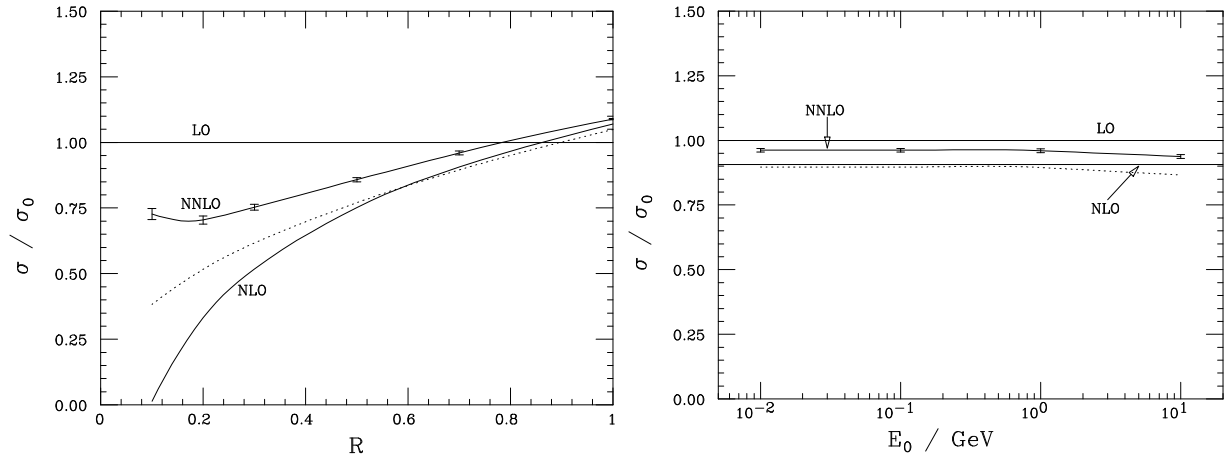

  \centerline{
    \resizebox{!}{6cm}{\includegraphics{profile_03.ps}}
    \hfill
    \resizebox{!}{6cm}{\includegraphics{profile_04.ps}}
    }
  \caption[]{{\it The radius dependence with $E_0=1$~GeV (left) and seed
      cell threshold dependence with $R=0.7$ (right) of the inclusive
      jet cross section in the improved iterative cone algorithm, in
      which midpoints of pairs of jets are used as additional seeds for
      the jet-finding, in fixed-order (solid) and all-orders (dotted)
      calculations.}}
  \label{improved}
\vspace*{-2ex}
\end{figure}

A simple solution to this problem was proposed some time ago\cite{LdP}.
It is a simple modification to the algorithm used in both the
theoretical calculation and the experimental measurement:
\begin{quote}
  \em After finding all possible jets using the seed cells, rerun the
  algorithm using the midpoint of all pairs of jets found in the first
  stage as additional seeds\footnote{To save computer time, it is
    sufficient to just do this for jet pairs that are between $R$ and
    $2R$ apart.}.
\end{quote}
This means that the results become insensitive to whether there was a
seed cell in the overlap region, and hence to the energy threshold
$E_0$.  Cross sections are well-behaved and calculable order by order in
perturbation theory, as shown in Fig.~\ref{improved}.
Experimental results would be little changed by this modification
(compare the all-orders results of Figs.~\ref{itxsec}
and~\ref{improved}), but the theoretical predictions would be enormously
improved (compare the NNLO results of Figs.~\ref{itxsec}
and~\ref{improved}).

It should be stressed that this does not completely remove the problem
of merging and splitting of overlapping cones.  It merely relegates it
to a procedural problem: one should state clearly the procedure one
uses, and apply it equivalently to theory and experiment.  Provided that
that procedure uses information from all the jets in a democratic way
(i.e.~not keeping track of which jets came from seed cells, and which
from the additional seeds), it will not spoil the improved properties of
the algorithm.

\begin{figure}
  \centerline{
    \resizebox{!}{6cm}{\includegraphics{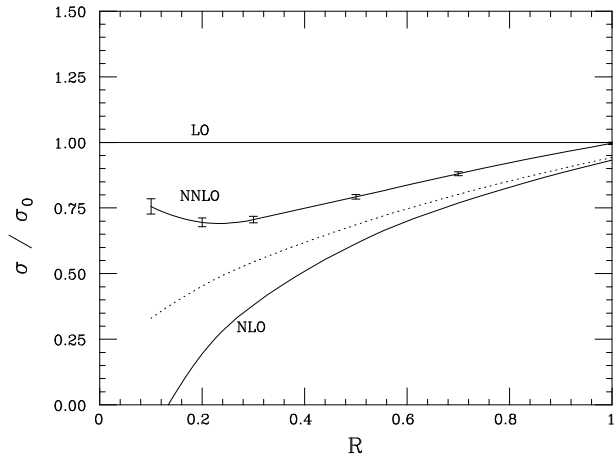}}
    }
  \caption[]{{\it The `radius' dependence of the inclusive jet cross
      section in the $\kt$ jet algorithm in fixed-order (solid) and
      all-orders (dotted) calculations.}}
  \label{kt}
\vspace*{-2ex}
\end{figure}

We finish this section by noting that using the $\kt$ algorithm removes
these problems completely.  It is fully infrared safe, and has no
overlap problem, because every final-state particle is assigned
unambiguously to one and only one jet.  We show results in
Fig.~\ref{kt}.
Unless there are factors of which we are unaware, abandoning the
iterative cone algorithm and using the $\kt$ algorithm instead would be
an even better solution than the previous one.

\section{The jet shape}

The jet shape is, at present, the most common way of resolving internal
jet structure.  It is inspired by the cone-type jet algorithm, but its
use is not restricted to cone jets.  It is defined by first running a
jet algorithm to find a jet axis.  The jet shape $\Psi(r;R)$ is then:
\begin{equation}
  \label{Psi}
  \Psi(r;R) =
  \frac{ \sum_i E_{Ti} \; \Theta(r-R_{i\mathrm{jet}}) }
       { \sum_i E_{Ti} \; \Theta(R-R_{i\mathrm{jet}}) },
\end{equation}
where the sum over $i$ can be over either all particles in the event, as
used by CDF and D\O, or only those particles assigned to the jet, as
used by ZEUS.  We have found that using cone-type jet definitions, there
is little difference between the two (less than 10\% even at the jet
edge).  However, if the jet is defined in the $\kt$ algorithm,
there are strong reasons for preferring the definition in which
the sum is only over those particles assigned to the jet.  For now, we
concentrate on the more commonly-used definition in which the sum is
over all particles.  Thus $\Psi$ is the fraction of all energy within a
cone of size $R$ around the jet axis that is within a smaller cone of
size $r$, also around the jet axis.  Clearly we have $\Psi(R;R)=1$, with
$\Psi(r;R)$ rising monotonically~in~$r$.

It is often more convenient to work in terms of the differential jet
shape:
\begin{equation}
  \label{psi}
  \psi(r;R) = \frac{d\Psi(r;R)}{dr}.
\end{equation}
Thus $\psi\,dr$ is the fraction of all energy within a cone of size $R$
around the jet axis that is within an annulus of radius $r$ and width
$dr$, centred on the jet axis.

The NLO matrix elements for
the jet cross section determine the jet shape at LO.
However, we can avoid having to use the virtual matrix elements, by noting
that they only contribute to $\psi(r;R)$ at exactly $r=0$.  Thus we
can calculate $\psi(r;R)$ for all $r>0$ from the tree level matrix
elements and then get the contribution at $r=0$ from the fact that it
must integrate to 1, i.e.
\begin{equation}
  \psi(r;R) = \delta(r) +
  \Bigl( \psi_{\mathrm{tree\;level}}(r;R) \Bigr)_+,
\end{equation}
where $f(r;R)_+$ is a distribution defined in terms of the function
$f(r;R)$ by $f(r;R)_+=f(r;R)$ for $r>0$ and $\int_0^Rf(r;R)_+dr=0$.
It is straightforward to integrate the tree-level matrix elements to
obtain the LO prediction for the jet shape.
However, it is also useful to have an analytical approximation to the
matrix
elements to work with.  This can be done using the modified leading
logarithmic approximation (MLLA), in which we have contributions from
soft and/or collinear final-state emission, and soft initial-state
emission.
For a quark jet we obtain\cite{paper}
\begin{equation}
  \psi_q(r) = \frac{C_F\as}{2\pi} \, \left[ \frac2r
  \left( 2\log\smfrac1Z - \smfrac32(1-Z)^2 \right) \right]_+
  +\psi_i(r),
\end{equation}
where
\begin{equation}
  Z=\left\{ \begin{array}{ll}
      \frac r{r+R}     & \mbox{if $r<(\Rsep-1)R$,} \\
      \frac r{\Rsep R} & \mbox{if $r>(\Rsep-1)R$}
    \end{array} \right.,
\end{equation}
and $\Rsep$ parametrizes the jet algorithm: $\Rsep=1$ in the iterative
cone and $\kt$ algorithms, $\Rsep=2$ in the improved cone algorithm.
For a gluon jet,
\begin{eqnarray}
  \psi_g(r) &=& \frac{C_A\as}{2\pi} \, \left[ \frac2r
  \left( 2\log\smfrac1Z - (1-Z)^2
    \left(\smfrac{11}6 - \smfrac13Z + \smfrac12Z^2\right) \right)
  \right]_+
\nonumber\\
  &+& \frac{T_RN_f\as}{2\pi} \, \left[ \frac2r
  (1-Z)^2\left(\smfrac23 - \smfrac23Z + Z^2\right) \right]_+
  +\psi_i(r),
\end{eqnarray}
where $N_f$ is the number of flavours.  The contribution $\psi_i(r)$
comes from initial-state radiation that is clustered into the jet, and
is the same for quark and gluon jets.  It is given by:
\begin{equation}
  \psi_i(r) = \frac{C\as}{2\pi} \, \left[ 2 r
    \left(\frac1{Z^2}-1\right) \right]_+,
\end{equation}
where $C$ is a factor that in principle depends on the kinematics and
colour flow of the hard scattering, but in practice is well approximated
by a constant, $C\sim C_F\sim C_A/2$, for which we use $C=C_A/2$ for all
numerical results.

\begin{figure}
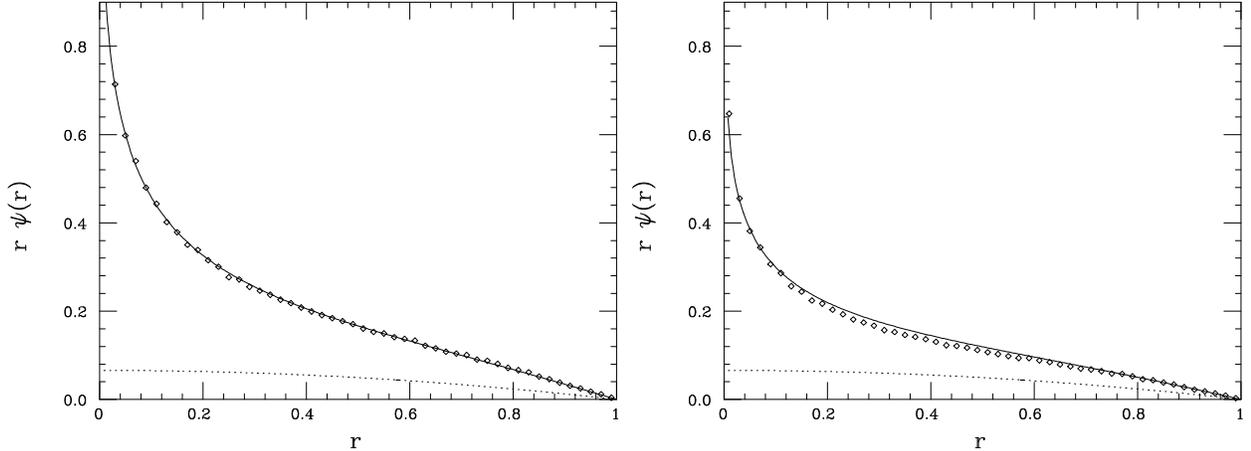

  \centerline{
    \resizebox{!}{6cm}{\includegraphics{profile_06.ps}}
    \hfill
    \resizebox{!}{6cm}{\includegraphics{profile_07.ps}}
    }
  \caption[]{{\it The jet shape at leading order in the $\kt$ algorithm
      for a 50~GeV jet at $\eta=0$ (left) and $\eta=3$ (right),
      according to the exact tree-level matrix elements (points) and our
      analytical formulae (curves).  The contributions of the
      initial-state component of the latter are shown separately as the
      dotted curves.}}
  \label{LO}
\vspace*{-2ex}
\end{figure}

In Fig.~\ref{LO} we show the results of both the full LO matrix element
integration, and our analytical approximation to it, for the $\kt$
algorithm.
As with all the numerical results in this paper, we use the CTEQ4M
parton distribution functions\cite{CTEQ}.  We see remarkably good
agreement between the full result and the analytical approximation.  The
contribution from initial-state radiation is shown separately, and is
clearly essential for this good agreement.

Having seen that the analytical results approximate the full LO matrix
element well, we move to higher orders to see how much we can improve
them.  We find that the jet shape in the iterative cone algorithm is
strongly dependent on the seed cell threshold, as anticipated from the
arguments of the previous section.  We neglect it from further
discussion.

\begin{figure}
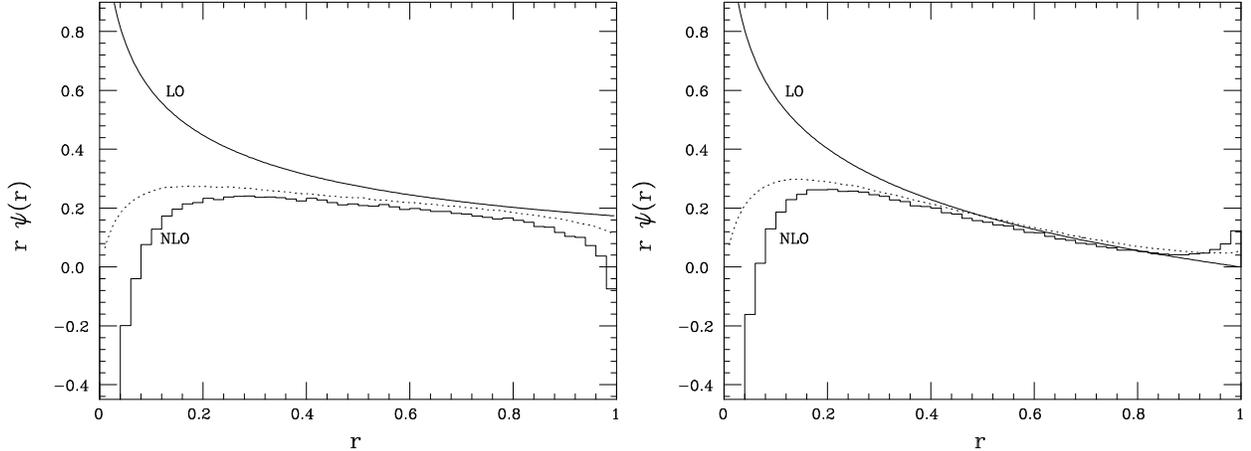

  \centerline{
    \resizebox{!}{6cm}{\includegraphics{profile_13.ps}}
    \hfill
    \resizebox{!}{6cm}{\includegraphics{profile_14.ps}}
    }
  \caption[]{{\it The jet shape in the improved cone (left) and $\kt$
      (right) jet algorithms in fixed-order (solid) and all-orders
      (dotted) calculations.}}
  \label{ktshape}
\vspace*{-2ex}
\end{figure}

In Fig.~\ref{ktshape} we show results in the improved cone algorithm and
the $\kt$ algorithm.
In the improved cone algorithm, the NLO corrections are rather large
(note that the normalization is outside the control of this
approximation, and we should look at the shape of the corrections only).
Close to the jet edge, they diverge to negative infinity, a typical
`Sudakov shoulder' effect\cite{CW}, analogous to the $C$-parameter
distribution in $e^+e^-$ annihilation for $C\sim\frac34$.  The
corresponding logarithms of $(R\!-\!r)$ must be resummed to all orders
for a reliable prediction.  The correction in both algorithms becomes
large and negative at small $r$ due to logarithmic terms in $r$, which
can be resummed to all orders in $\as$ to give a physically-behaved
prediction\cite{paper}.

In the $\kt$ algorithm, the NLO corrections diverge to positive infinity
near the edge of the jet.  As discussed in \cite{paper}, this is an
artifact of the fact that we define the jet shape
using all particles in the event.  If we instead use only those
particles assigned to the jet, we obtain a NLO result that is continuous
at $r\!=\!R$.
To avoid these large higher order terms, we recommend that in future the
jet shape be defined using \emph{only those particles assigned to the
  jet by the jet algorithm}.

\begin{figure}
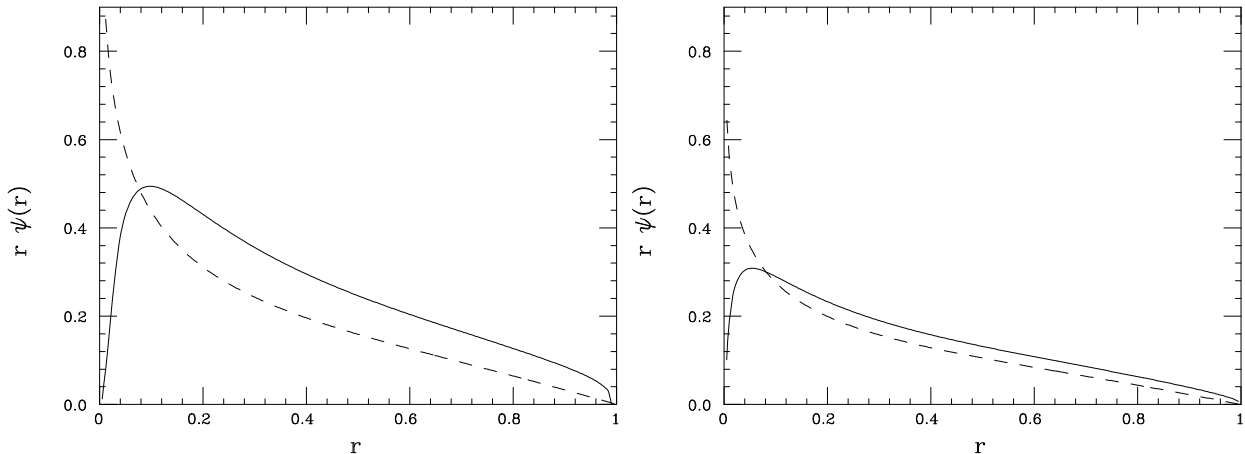

  \centerline{
    \resizebox{!}{6cm}{\includegraphics{profile_22.ps}}
    \hfill
    \resizebox{!}{6cm}{\includegraphics{profile_26.ps}}
    }
  \caption[]{{\it Total effect of running coupling, power corrections
      and resummation on the shape of a 50~GeV (left) and 250~GeV
      (right) jet in the $\kt$ algorithm: LO (dashed) and with
      everything (solid).}}
  \label{final}
\vspace*{-2ex}
\end{figure}

A formalism has been developed over many years for summing various
logarithmically-enhanced terms to all orders in $\as$.  In doing so, one
inevitably ends up integrating over the Landau pole of the running
coupling in perturbative calculations, signalling that non-perturbative
confinement effects play a crucial r\^ole.  In the Dokshitzer-Webber
approach\cite{DW1}, these enter through {\it a priori} unknown, but
universal, constants.  The same one determines the jet shape as the
average value of several event shapes in $e^+e^-$ annihilation, like
thrust.  Thus jet shapes offer an excellent opportunity to test this
universality, by comparing the quark-dominated jets of $e^+e^-$
annihilation with the gluon-dominated jets of hadron collisions.  More
details can be found in \cite{paper}.  The net result of these
corrections is shown in Fig.~\ref{final}.
At $E_T=50$~GeV, they roughly double the amount of energy near the edge
of the jet.  Even at high transverse energy, $E_T=250$~GeV, it is
increased by about 50\%, although most of this is accounted for by
running coupling effects.

Looking at Figs.~\ref{ktshape} and~\ref{final}, one sees that neglected
NLO, logarithmically enhanced, and power-suppressed terms are all very
important in determining the jet shape distribution.  We should not
therefore be terribly surprised if LO predictions do not fit data very
well.  Quantitative studies of internal jet structure will only be
possible once the long-awaited NLO corrections have been calculated.

\section{Subjet structure of jets}

The jet shape is largely inspired by cone-type jet algorithms, although
it can be studied in cluster algorithms.  In cluster algorithms however,
it may seem more natural to study internal jet structure by using the
same clustering algorithm, but stopping before the jet is complete, to
define {\em subjets}.  This is much more closely related to how we think
that jets evolve: not by a gradual spreading in angle, but by
iteratively splitting into subjets, subsubjets and so on, until
eventually splitting to hadronic resonances and thence to stable
hadrons.

Within the cluster algorithm, subjets can be defined by keeping track of
which particles ended up in the jet we are interested in, and rerunning
the clustering algorithm using only those particles.  The clustering is
stopped when all the $d_{ij}$ are above some cutoff:
\begin{equation}
  y_{ij} \equiv \frac{d_{ij}}{E_{T\mathrm{jet}}^2} > \ycut.
\end{equation}
This is very similar to the way in which quark jets are studied in
$e^+e^-$ annihilation.

\begin{figure}
  \begin{minipage}[t]{0.55\linewidth}
    \centerline{
      \resizebox{!}{6cm}{\includegraphics{lathuile_01.ps}}
      }
    \caption[]{{\it Leading order (dashed) and resummed results for the
        subjet multiplicity in a 100~GeV jet at the Tevatron.}}
    \label{subjet1}
  \end{minipage}\hfill\begin{minipage}[t]{0.4\linewidth}
    \centerline{
      \resizebox{!}{6cm}{\includegraphics{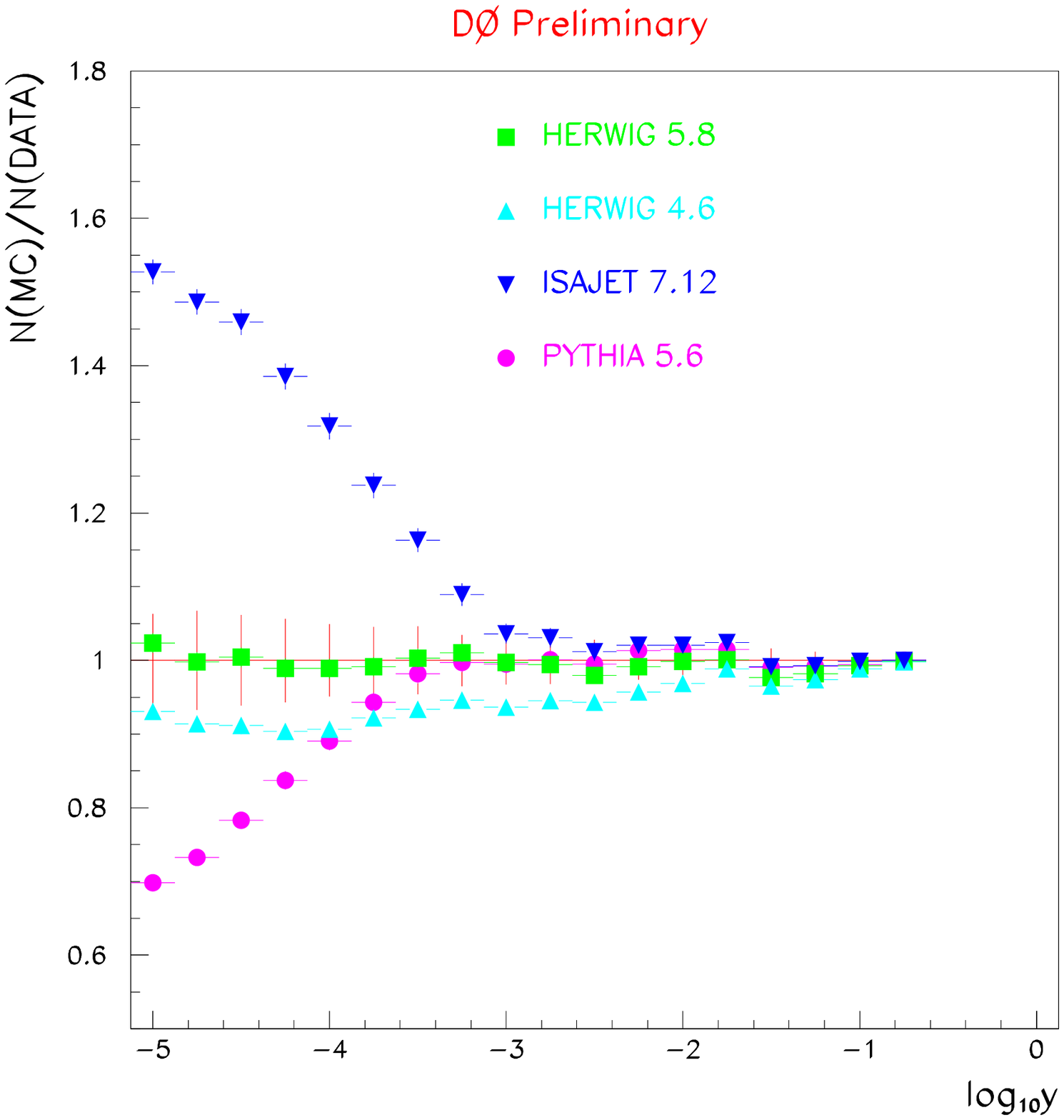}}
      }
    \caption[]{{\it The subjet multiplicity in a 300~GeV jet at detector
        level in various models, normalized to the data.}}
    \label{subjet2}
  \end{minipage}
\vspace*{-2ex}
\end{figure}

The simplest quantity one can imagine studying is the average number of
subjets as a function of $\ycut$.  As discussed in \cite{S}, when
$\ycut$ is small, the perturbative expansion is spoiled by large terms
that arise at every order, $\as^n\log^{2n}\ycut$, and these must be
summed to all orders for a reliable prediction.  These come from
final-state emission, and are identical to those in $e^+e^-$
annihilation, as are part of the next-to-leading correction,
$\as^n\log^{2n-1}\ycut$.  However, at this level, initial-state
radiation also contributes, making the results differ from the $e^+e^-$
annihilation case.  These terms can also be resummed, giving a
prediction that is uniformly reliable for all $\ycut$\cite{S}.  The
results are shown in Fig.~\ref{subjet1}.
We see that the resummation is essential for small $\ycut$, and that the
inclusion of the initial-state terms only makes a relatively small
difference.  In Fig.~\ref{subjet2}, we show the experimental results for
a 300~GeV jet at detector level in comparison with various Monte Carlo
event generators\cite{A}.  The agreement is remarkable, at least with
models that include a full account of colour coherence\cite{HW}.  At the
left-hand side of the plot we are probing 300~GeV jets at a scale of
only 1~GeV.  More recent data can be found in \cite{Sn,L}.

Once we have defined and counted the subjets, we can probe their
distribution relative to the jet axis in just the same way as one
normally does with the particles in a jet.  The cutoff, $\ycut$, can be
tuned to choose to sit in a fully perturbative regime (large $\ycut$),
or a fully hadronized regime (for $\ycut\to0$ every hadron is considered
a subjet of its own).  That way we can start in a well-understood
perturbative region, and gradually switch on hadronization in a
controlled way.  For example one can define the jet shape using subjets
rather than particles, and one finds almost negligible parton$\to$hadron
corrections for reasonable $\ycut$ values\cite{Sn,L}.

\section{Summary}

In recent years there has been a growth of interest in the internal
structure of jets.  This is being given added impetus at the moment by
the fact that a full NLO calculation is expected soon.  By studying
jets' internal structure, we are able to learn a great deal about the
process by which hard partons are confined into jets of hadrons.

This renewed interest, and the ever-increasing accuracy of theoretical
calculations, has prompted a critical evaluation of the quality of jet
definitions in current use.  We have found that they are insufficient
for the level of accuracy required, and should be improved as described
above, or better still replaced by cluster-type definitions like the
$\kt$ algorithm.

Higher order corrections, all-orders resummation and non-perturbative
hadronization corrections are all expected to be important in
determining the jet shape, and we should not be surprised if LO
calculations do not describe the data well.

Subjet studies offer a much greater degree of flexibility than the jet
shape alone.  One can choose to work in a highly-perturbative regime,
which should be an ideal place to measure $\as$ once we have NLO
calculations, since most dependence on the absolute normalization and
parton distribution functions drops out.  Or one can choose to lower
$\ycut$ and study the onset of hadronization, eventually ending up at
the usual hadronic final state.  In this case, one has the confidence of
knowing that the perturbative `background' is under good control, and
can ascribe deviations to the non-perturbative hadronization process.

It is to be hoped that with the NLO calculation to hand, and a greater
degree of dialogue between theorists and experimenters, a new generation
of internal jet measurements will emerge, shedding new light on the
nature of jets and confinement.

\end{document}